\documentstyle[aps,multicol,floats,epsfig]{revtex}
\begin{document}

\draft

\title{Formation and manipulation of a metallic wire of single gold atoms}
\author{A.I. Yanson$^{\star }$, G. Rubio Bollinger$^{\dagger }$,
H.E. van den Brom$^{\star }$, N. Agra\"{\i}t$^{\dagger }$, J.M. van
Ruitenbeek$^{\star }$}
\address{$^{\star }$Kamerlingh Onnes Laboratorium, Leiden University, PO Box 9506, NL-2300
RA Leiden, The Netherlands\\ $^{\dagger }$Laboratorio de Bajas
Temperaturas, Dept. F\'{\i}sica de la Materia Condensada C-III,
Instituto Universitario de Ciencia de Materiales ``Nicol\'{a}s
Cabrera'', Universidad Aut\'{o}noma de Madrid, E-28049 Madrid,
Spain}
\maketitle

\begin{abstract}
The continuing miniaturization of microelectronics raises the prospect of nanometre-scale devices with mechanical and electrical properties that are qualitatively different from those at larger dimensions. The investigation of these properties, and particularly the increasing influence of quantum effects on electron transport, has therefore attracted much interest. Quantum properties of the conductance can be observed when `breaking' a metallic contact: as two metal electrodes in contact with each other are slowly retracted, the contact area undergoes structural rearrangements until it consists in its final stages of only a few bridging atoms \cite{Jan NATO,Sutton,Landman}. Just before the abrubt transition to tunneling occurs, the electrical conductance through a monovalent metal contact is always close to a value of $2e^{2}/h (\approx (12.9 \textrm{k}\Omega )^{-1})$, where $e$ is the charge on an electron and $h$ is Plack's constant \cite{Agrait,Pascual,Krans}. This value corresponds to one quantum unit of conductance, thus indicating that the `neck' of the contact consists of a single atom \cite{Nature2}. In contrast to previous observations of only single-atom necks, here we describe the breaking of atomic-scale gold contacts, which leads to the formation of gold chains one atom thick and at least four atoms long. Once we start to pull out a chain, the conductance never exceeds $2e^2/h$, confirming that it acts as a 
one-dimensional quantized nanowire. Given their high stability and the ability to support ballistic electron transport, these structures seem well suited for the investigation of atomic-scale electronics.
\end{abstract}


Previous studies on metallic contacts of atomic dimensions have shown remarkable properties of such structures, including conductance quantisation and superior mechanical strength compared to the bulk. Experimental techniques, most common being scanning tunnelling microscopy (STM) and mechanically controllable break-junctions (MCB), are all based on piezoelectric transducers which allow fine positioning of two metal electrodes with respect to each other. STM, in which the tip is driven into contact with a metal surface and the conductance is measured during subsequent retraction, has been widely used for this purpose \cite{Agrait,Pascual,Brandbyge,Rubio}. In the alternative method of
MCB one starts with a macroscopic notched wire \cite{Muller}, or a nanofabricated metal bridge \cite{Ruitenbeek} mounted on a flexible substrate. The wire (or bridge) is broken at low temperatures in vacuum, and contact is re-established between the fracture surfaces by piezoelectric control of substrate bending. In this work we have used both MCB and a very stable STM at liquid helium temperatures to produce and study chains of single gold atoms. In each case, high purity ($99.99+\%$) gold was used. Conductance was measured at a 10\,mV DC voltage bias with 1\% accuracy. 
\begin{figure}[!t]
\begin{center}
\leavevmode
\epsfig{figure=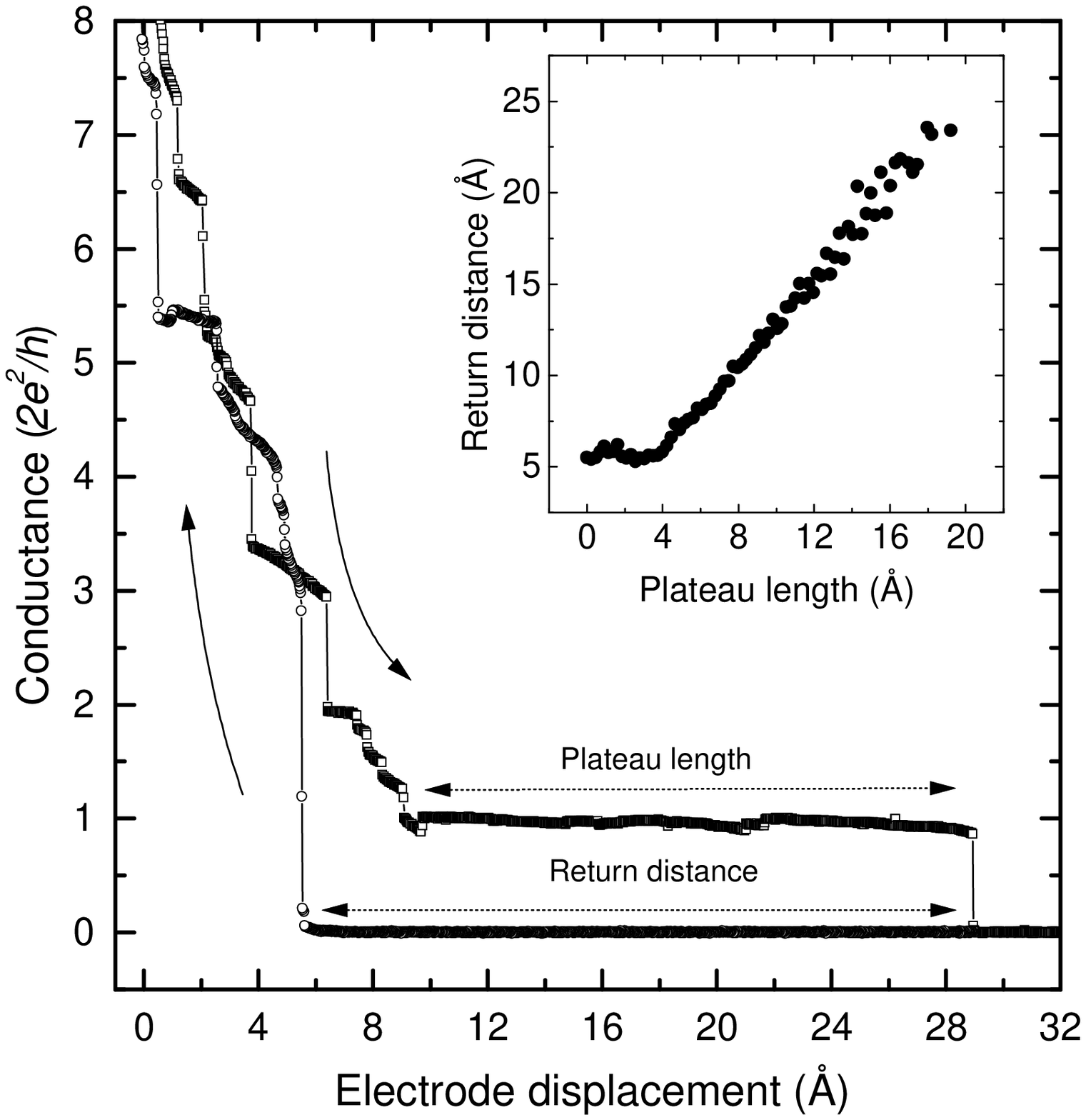,width=12cm}
\end{center}
\caption{The conductance as a function of the displacement of the two gold electrodes with respect to each other in an MCB experiment at 4.2\,K. The trace starts at the upper left, coming from higher conductance values (squares). A long plateau with a conductance near 1\,G$_{0}$ is observed and after a jump to tunnelling we need to return by a little more than the length of the long plateau to come back into contact (circles). The inset shows the average of the return distance as a function of the length of the long plateau (recorded with an STM at 4.2 K). The relation is approximately 1:1, with an offset of 5\,\AA\, which is probably due to the elasticity of the atomic structure.}
\label{fig:fig1} 
\end{figure}

An example of a conductance curve obtained while stretching a gold nanocontact is presented in Fig. 1. The curve reflects the evolution of some particular atomic configuration, during which the conductance decreases in a series of sharp vertically descending steps, with a gradual slope on the plateaux in between. These steps have been shown to be the result of atomic structural rearrangements \cite{Rubio,Todorov}. The curves for successive rupture sequences do not repeat in detail, since they depend on the exact atomic positions in the contact. However, many curves have one striking feature in common, being the remarkable length of the last conductance plateau just before rupture, at the value of about one conductance quantum, $G_0=2e^2/h$. Previous experiments have shown that the conductance at the last plateau for monovalent metals is usually found close to 1\,$G_0$ (refs\,\onlinecite{Agrait,Pascual,Krans}), and it has been argued that these plateaux correspond to a contact with a single atom at the narrowest cross section \cite{Nature2}. Examples have been found previously of exceptionally long stretched last plateaux, in particular for gold (see, for example, ref.\ \onlinecite{thesis}), but this has not received due attention. The interest becomes clear when we recognise that the conductance of a point contact is determined predominantly by the size of its narrowest cross-section and that, during contact elongation, all structural transformations are localised to the neck region \cite{Sutton,Landman,Todorov}. Fig.\,1 shows that during the last stage of elongation the conductance stays in a limited range of values corresponding to one atom in cross section, while the contact is being stretched over distances up to 20\,\AA\ . This suggests the possibility that the contact stretches to form a chain of single atoms. Since we cannot image the atomic structure of the neck directly, we performed the following experiments in order to test this hypothesis.

Fig.\,1 illustrates that the distance one needs to travel back in order to    re-establish contact after rupture is almost equal to the length of the last plateau itself. We have recorded these lengths for a large number of contacts and the average return distance is plotted in the inset in Fig.\,1 as a function of the length of the last plateau. The relation is approximately 1:1, suggesting that a fragile structure is formed with a length corresponding to that of the last plateau, which is unable to support itself when it breaks and collapses onto the banks on either side.

The probability distribution for pulling a long last plateau as a function of its length was obtained by recording a histogram of plateau lengths, and is given in Fig.\,2. Instead of a smooth distribution we find a series of three equidistant peaks, with a shoulder at the fourth and fifth position. The probability of pulling a structure of length $L$ decreases rapidly for large $L$ and drops below $10^{-4}$ for $L>20$\,\AA, so the longest observed plateaux have a length of 25\,\AA. The shape of the histogram suggests that the nanobridges prefer to be elongated by integer multiples of the (somewhat stretched) Au-Au inter-atomic distance. 
\begin{figure}[!t]
\begin{center}
\leavevmode
\epsfig{figure=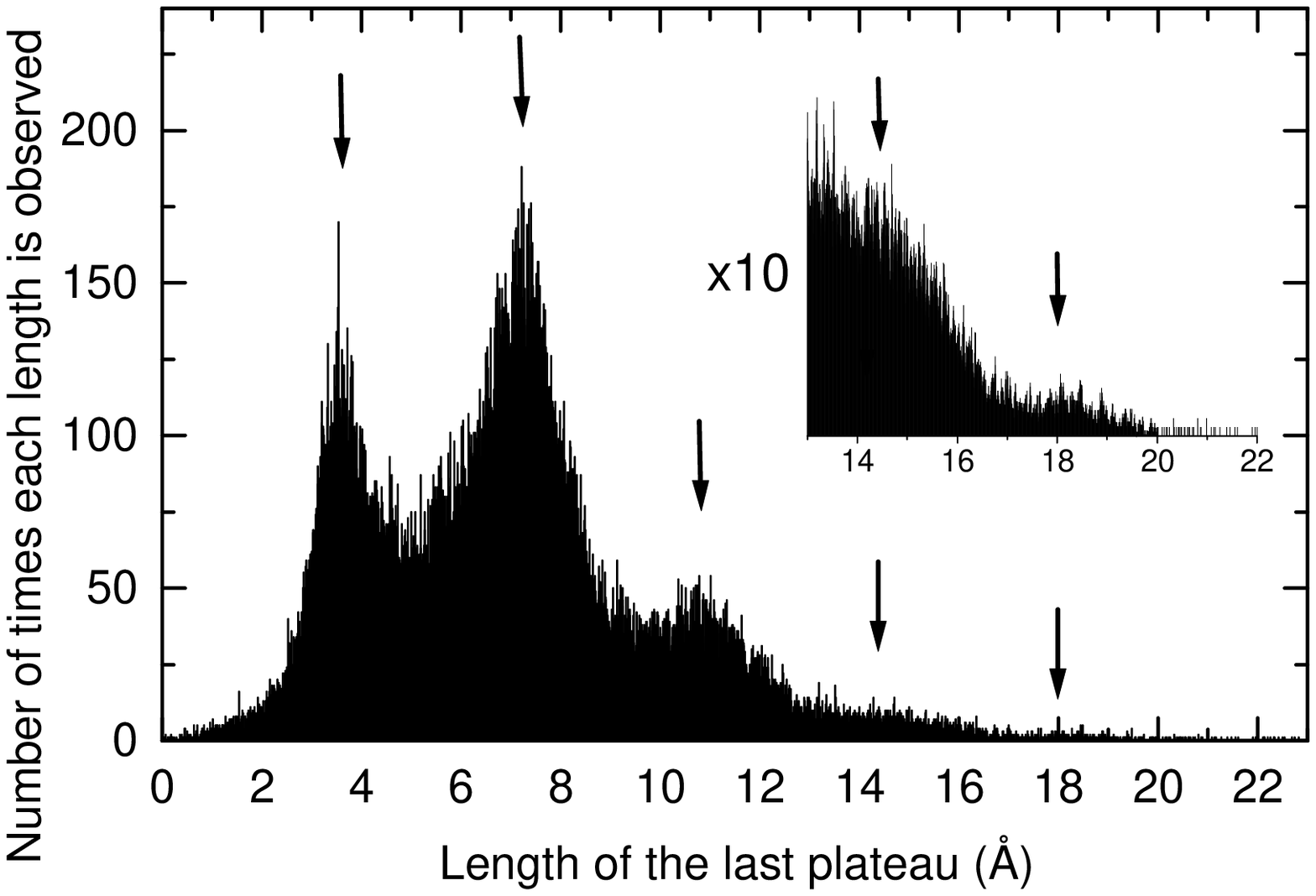,width=12cm}
\end{center}
\caption{The distribution of lengths for the last plateau, obtained in
$\sim 100,000$ experiments similar to those described in Fig. 1, shows
a number of equidistant maxima. The arrows are positioned at multiples of
3.6\,\AA . The data were recorded with an MCB at 4.2\,K. The length of the last plateau was defined as the distance between the points at which the conductance drops below 1.2\,$G_{0}$ and 0.8\,$G_{0}$, respectively. The inset shows the tail of the distribution on a 10$\times $ expanded scale. The accuracy for the calibration of the length scale is 30\%.}
\label{fig:fig2} 
\end{figure}

As a further test, upon formation of a plateau of a given length, one of the electrodes was moved laterally. For a short length of the last plateau (Fig.\,3a), when we start oscillating the STM tip sideways (Fig.\,3b), we observe rupture of the contact at low oscillation amplitudes. Repeating the experiment for longer elongation of the last plateau (Fig.\,3c), fracture is observed at much larger lateral displacement amplitudes (Fig.\,3d) and the bridge is found to collapse. The small jumps in the conductance in Fig.\,3b and 3d have the same origin as those found during elongation of the last plateau in Fig.\,1. That is, they are due to atomic structural rearrangements, which have only a minor effect on the conductance as long as the narrowest cross section is still a single atom. The two atomic configurations at each side of the small conductance jump are separated by an energy barrier, which can be surmounted by applying a sufficiently large force. This force stretches the atomic bridge over a length proportional to its elasticity. Consequently, the hysteresis of the conductance jumps that occur as the structure is moved laterally (Fig. 3d) indicates that longer structures have a larger lateral elasticity than short ones.
 
We find that there is a significant probability for formation of a long bridge with a length given by that of the last plateau, which completely collapses upon breaking. Its smallest cross section is one atom and it breaks at multiples of approximately 3.6\,\AA\  in length, which would correspond to a stretched Au-Au bond distance. We can swing the bridge sideways by a distance comparable to its length. We conclude that all the evidence combines to show that we are pulling, atom by atom, a freely suspended chain of single Au atoms. The peaked structure of Fig.\,2 clearly marks the observation of 4 or 5 atoms long chains. Although we cannot resolve the peak structure for longer plateaux, extrapolating the periodic structure suggests that the longest plateaux observed correspond to chains of 7 atoms.

Despite its low probability of formation, once an atomic chain is pulled it remains very stable: some of the longest ones obtainable in our experiments have been held stable for as long as 1 hour, after which we stopped the experiment. They can sustain enormous current densities of up to $8\times 10^{14}$ A$/$m$^{2}$ (currents up to 80 $\mu $A or a voltage up to 1 V), proving that the electron transport is ballistic, and most of the power is dissipated in the electrodes far away from the contact. This makes them suitable candidates as conductors in research of atomic electronic circuits.
\begin{figure}[!t]
\begin{center}
\leavevmode
\epsfig{figure=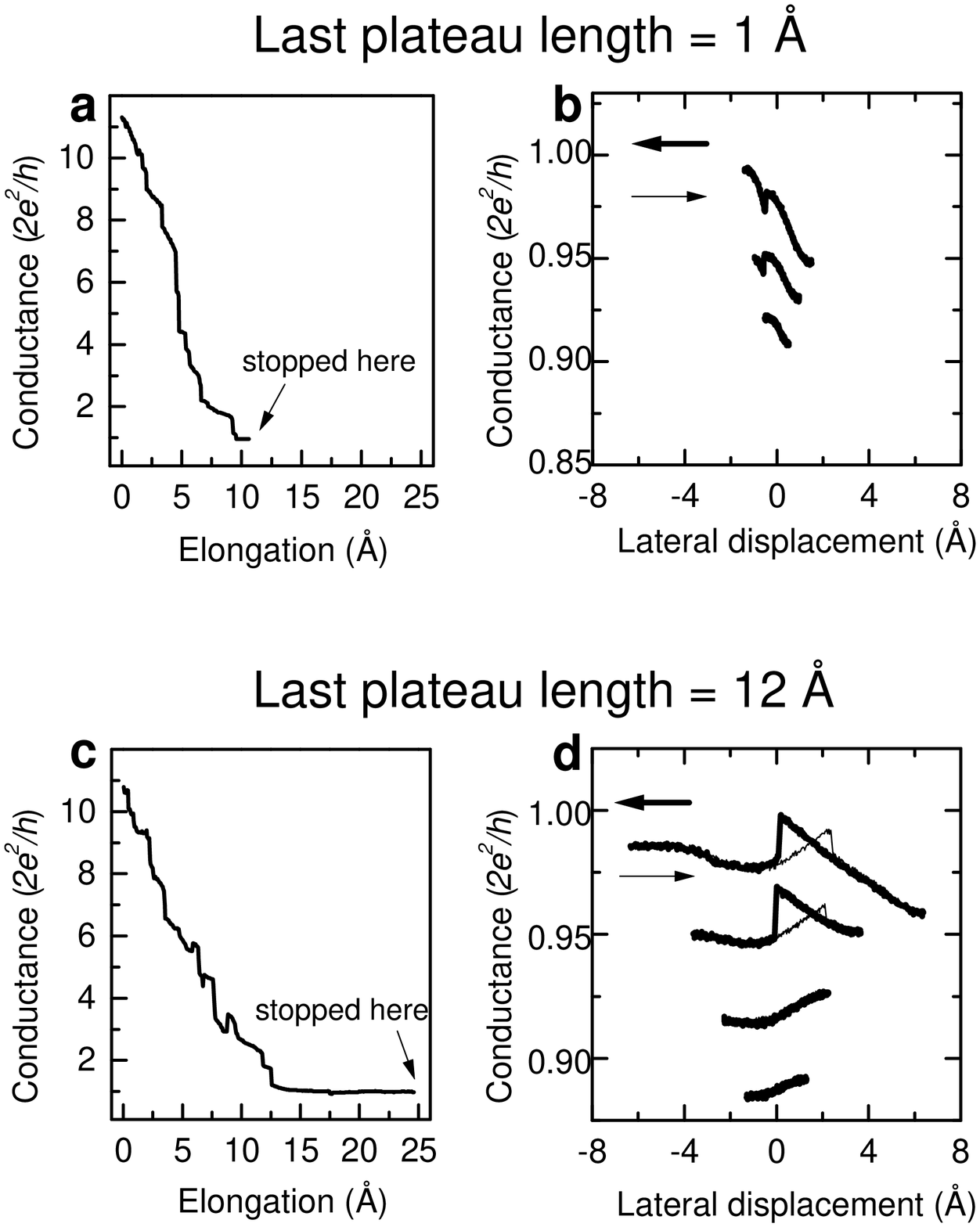,width=12cm}
\end{center}
\caption{Swinging an atomic contact sideways by lateral displacement of one of its ends in an STM experiment. First, the elongation of the last plateau is stopped early (\textbf{a}), at a length of about 1\,\AA\ (marked with an arrow). Then, one end of the contact is cyclically displaced perpendicular to the contact axis (\textbf{b}) while slowly increasing the displacement amplitude. The curves for smaller swing amplitudes are shifted by increments of $-0.03\ G_{0}$ for clarity. The contact breaks at a lateral displacement of 1.4\,\AA\ . Performing the same experiment for a last plateau of 12\,\AA\ length (\textbf{c}) gives quite different results: rupture occurs at much larger lateral swing amplitudes of 6.3\,\AA\ (\textbf{d}) and the contact bridge collapses. For small swing amplitudes the contact behaves elastically and the conductance for both directions of the swing superpose. As the amplitude is increased small hysteretic jumps in conductance take place (note the expanded scale), resulting from slight rearrangements of the atoms in the chain or at its bases.}
\label{fig:fig3} 
\end{figure}

Once the conductance drops below 1 $G_{0}$ at the beginning of the chain formation, it never rises above this value. This is consistent with the notion of the conductance being perfectly described by one single quantum mode \cite{Nature2}. Fluctuations to lower values correspond to a reduced transmission probability for this mode as a result of back-scattering, but the maximum value should be limited to 1 $G_{0}$, in agreement with the observations. In this sense, our metallic atomic wire is a perfectly quantized one-dimensional conductor. Previously Yazdani {\it et al.}\cite{Eigler} succeeded in measuring the conductance of a chain of two atoms of the noble gas xenon, which was found to be orders of magnitude lower than the quantum unit of conductance due to the lack of metallic bonding. A recent calculation by Lang\cite{Lang} on the single atomic chains of up to 4 sodium atoms predicts their conductance to fall below $0.7 G_0$, fluctuating with the chain length. This is somewhat smaller than the experimental values, which almost all fall in the range from 0.8 to 1 $G_0$ (see, for example, Figs 1 and 3). This discrepancy may be influenced by the interface to the `jellium' electrodes used in the calculation.

Recent molecular dynamics simulations of metallic nanocontacts by Sutton and Todorov (unpublished), S\o rensen \textit{et al}. \cite{Sorensen,Sorensen2} and Finbow \textit{et al.} \cite{Finbow} support the plausibility of occasional formation of atomic chains, although the authors note that their model inter-atomic potential may not be reliable for these unusual one-dimensional structures. Recently, advanced 
first-principle molecular dynamics calculations have been performed for Na nanocontacts at 190\,K \cite{Barnett}, which showed no indication for chain formation. Although it is not clear to us at this moment what determines the difference in behaviour between gold and sodium, this result is consistent with the fact that we do not observe chain formation for sodium in the low-temperature experiments. 

The atomic chain structures open many avenues for further investigations. Studying the mechanical properties of this unusual form of matter should enable stringent tests of our understanding of 
interatomic potentials. One may search for evidence of one-dimensional excitations such as phonons and plasmons. Further, the electronic properties of these chains are expected to be those of a perfect one-dimensional conductor, where the electron-phonon interaction may lead to a Peierls transition. The electron-electron interaction may lead to the formation of a so-called Tomonaga-Luttinger liquid \cite{fisher}, which replaces the familiar Fermi liquid description for bulk metals when the conductor becomes 
one-dimensional.

\bigskip
\bigskip
\noindent\textbf{Acknowledgements}. We thank B. Ludoph for many
  discussions, S. Vieira and L.J. de Jongh for discussions and continuous     
  support and A.P. Sutton, T.N. Todorov, M.R. S\o rensen, M. Brandbyge and
  K.W. Jacobsen for communicating their results prior to
  publication. A.I.Y., H.E.v.d.B., J.M.v.R. were supported by FOM; N.A. and
  G.R.B. were supported by the CICYT. 

\end{document}